\documentclass[letterpaper,twocolumn,10pt]{article}
\usepackage{usenix2019,epsfig,endnotes}
\begin{document}


\title{Thinkey: A Scalable Blockchain Architecture }
\author[2]{Shan Chen}
\author[1]{Weiguo Dai}
\author[1]{Yuanxi Dai}
\author[1]{Hao Fu
\thanks{Corresponding author: fuhaots2009@qq.com}
}
\author[1]{Yang Gao}
\author[1]{Jianqi Guo}
\author[1]{Haoqing He}
\author[3]{Yuhong Liu
\thanks{Authors are sorted in alphabetical order.}
}
\affil[1]{Beijing Thinkey Science and Technology Ltd., Bejing, China}
\affil[2]{Georgia Institute of Technology, Atlanta, Georgia}
\affil[3]{Santa Clara University, Santa Clara, California}

\maketitle

\thispagestyle{empty}

\begin{abstract}
This paper presents Thinkey, an efficient, secure, infinitely scalable and decentralized blockchain architecture. It ensures system correctness and liveness by a multi-layer structure. 
In particular, the system is based on a double-chain architecture and uses a multi-layer consensus protocol to guarantee consistency. 
Thinkey also uses a novel account model which is based on Actor Model to support the complex logic in the multi-chain structure.
Experiment results show that the proposed Thinkey architecture can achieve higher throughput as the number of nodes increases.
\end{abstract}

\section{Introduction}
While the centralized Internet has promoted 
the rapid development of cyber systems and their communications, it also introduces many problems such as data ownership, information opaqueness, monopoly and lack of trust.
Today, people are increasingly looking forward to trustworthy public computing infrastructures, in which an entity's behavior is always carried out in an expected way. Such trustworthy infrastructures usually contain trustworthy hardware, systems, and software.

In such a context, a single entity can no longer meet people's need for trust. People started to create trustworthy distributed computing platforms which involve participation from individual computing nodes (\eg Bitcoin \cite{nakamoto2008bitcoin}) but are not controlled by any committee or individual.
 These computing nodes should be in the hands of diverse parties or individual clients running a program that is completely open-source with specifications open to everyone.

Blockchain is a \emph{trusted distributed computing platform} that fulfills the above requirements. 
Recently, studies on blockchain related technologies are gaining popularity and attracting attention from researchers in different fields, such as cryptography, distributed system, network, economy, and social science.
This field was originated from the work of Nakamoto \cite{nakamoto2008bitcoin},  who proposed the first blockchain protocol as Bitcoin,
and extended a lot of research directions, including consensus, privacy computing, mechanism design, network and data storage.
It also found numerous applications in many areas, such as financial systems, supply chains, health cares.

As a distributed platform that is composed of thousands of individual computing nodes without any central controller, blockchain systems are facing many challenges. One of the most critical challenges is to achieve permissionless consensus among the distributed computing nodes in an efficient and scalable way. This is particularly true in the scenario where the system continuously grows with an increasing amount of clients and data, which requires (1) high efficiency in terms of handling multiple concurrent transactions in parallel, and (2) scalability in terms of maintaining high efficiency when the system scales up. However, the performance of current blockchain platforms is still far away from being satisfactory.   
For example, the average transaction rates in Bitcoin have been limited to below 10 transactions per second (Tx/s). More importantly, when the processing limit is reached, the blockchain platforms may have to continue by discarding some of their clients' transactions, which will undermine the systems' availability and significantly increase the usage costs.

To improve the system efficiency and scalability to support large-scale transactions (\eg smart contract transactions), the sharding approach has been proposed as a feasible solution, which aims to divide transactions into multiple packages that can be handled in parallel by different committees of participating nodes \cite{corbett2013spanner}.  
The idea of network and transaction sharding was first proposed in Elastico \cite{luu2016secure} and promoted by Zilliqa \cite{zilliqa2017}, OmniLedger \cite{kokoris2018omniledger}, Chainspace \cite{al2017chainspace} and RapidChain \cite{zamani2018rapidchain}.  
However, current sharding solutions are still at their early stage and cannot well handle the following two main challenges:
(1) how to split the state into multiple shards and support cross-shard transactions efficiently; 
and (2) how to select nodes to ensure the trustworthiness of each committee.   
\liu{Based on my understanding, I thought shards and tasks mean the same thing and revise this paragraph accordingly. If my understanding is wrong, please feel free to revise the draft accordingly.
The major issue is to organize the order of the four challenges in a logical way.}

\subsection{Overview of Thinkey}
In order to solve the challenges discussed above, we have proposed and implemented Thinkey, a novel blockchain infrastructure that is fully decentralized, trusted and scalable.
In particular, our structure is based on a  double-chain architecture consisting of a {\em root chain} and {\em transaction chains}. 
The workloads are partitioned and handled by parallel transaction chains.
The root chain acts as the leader and coordinator of the entire system. 
In the transaction chain, we package the transactions into blocks and use the Byzantine consensus algorithm to reach consensus. 
The structure has great flexibility and scalability and can be dynamically adjusted so that each chain does not become a performance bottleneck of the entire system. In addition, as the number of chains increases, the throughput of the entire system increases linearly without generating too many redundant messages.
 
In addition, to efficiently handle cross-chain transactions, we propose a message-driven protocol based on Actor Model \cite{hewitt1973universal} in this work.
In particular, as the state of the entire network is partitioned by chains, and each chain is responsible for its own piece, a transaction involving multiple chains will appear with high probability as the number of chains increases. 
As a result, it requires efficient solutions to handle cross-chain transactions. In this work, we apply the Actor Model to our system, which allows implementing complex logic in an asynchronous and lock-free manner to keep chains concurrent and fully utilized.
In this model, each account represents an actor. 
We decouple a transaction which involves a set of accounts into multiple steps by messages. 
Each message will be sent to a unique recipient and will be received and executed by the corresponding chain.
Each chain is incentivized to handle messages to ensure the atomicity eventually. 

\smallskip
\noindent{\bf Our Contribution.} Particularly, this paper mainly makes the following contributions.
\begin{enumerate}
\item A four-layer encapsulation system blockchain framework has been proposed and implemented. In particular, the proposed layered architecture enables the abstraction and separation of diverse system functionalities and allows each layer to focus on its own functionality. The design of the proposed blockchain framework strengthens the blockchain system’s compatibility and extensibility, providing a new way for blockchain framework’s development.
\item A double-layer chain practical architecture that partitions transactions and states into parallel chains is proposed. The structure has great flexibility and scalability and can be dynamically adjusted, such that each chain does not become a performance bottleneck of the entire system.  
\item A new model of accounts in the multi-chain structure is proposed to process cross-chain transactions in an efficient and scalable way.
\end{enumerate}

\subsection{Related Work} \label{sec:related}
The blockchain technology has been successfully proved by Bitcoin \cite{nakamoto2008bitcoin} and Ethereum \cite{wood2014ethereum} through several years of operations, and shows great potential to be applied in various applications. 
However, the performance and scalability issues remain as the main obstacles that hinder the rapid development of blockchain systems. Researchers from different domains are conducting various research projects to improve blockchain performances and scalability. 
For example,
Bitcoin-NG \cite{eyal2016bitcoin} scales Bitcoin without changing the consensus protocol. It uses the same consensus protocol to elect a leader and commits
 ByzCoin \cite{kogias2016enhancing} extends Bitcoin-NG to achieve high transaction throughput through a variant of PBFT \cite{castro1999practical}.
Thunderella \cite{pass2018thunderella} has designed two chains to quickly confirm in optimistic situations.
By replacing the proof-of-work with the proof-of-state, the committee can be chosen based on their stakes \cite{gilad2017algorand,daian2017snow}. Conflux \cite{li2018scaling} uses the structure of the DAG to generate blocks at the same time. However, these algorithms still require all nodes to validate and process all transactions.

Elastico \cite{luu2016secure} was the first candidate for a secure sharding protocol for permissionless blockchain, where nodes are assigned to committees through proof-of-work and each committee is responsible for managing a subset of transactions (shard) consistently through PBFT \cite{castro1999practical}. A final committee collates sets of transactions received from committees into a final block and then broadcasts it. 
In order to meet the performance, the size of its committee needs to be small, resulting in a high probability of failure.
Zilliqa \cite{zilliqa2017} has improved the election protocol and consensus protocol for improved security.
OmniLedger \cite{kokoris2018omniledger} introduces a cross-shard commit protocol that atomically handles transactions affecting multiple shards. Chainspace \cite{al2017chainspace} improved its solution so that there are no clients involvement and support for smart contracts.
A recent study RapidChain \cite{zamani2018rapidchain} further optimizes these sharding protocols. It is resilient to Byzantine faults from up to a $1/4$ fraction.

\section{Blockchain: A New Generation of Computing}\label{sec:back}
Blockchain is a \emph{trusted distributed computing platform} that can act as a shared computational resource. Clients can view this platform as a new generation of computing facilities, to which they can send any requests (\ie transactions) to it. 
When the platform receives a request, it checks the legality of the request first, runs the request on some nodes, and returns the results to the client. Through this process, data and messages are transmitted on the platform to manage computing nodes and resources in a distributed way. In particular, a valid blockchain design should have the following properties. (a) {\em Safety:} the result is correct.
(b) {\em Liveness:} each valid request is processed within some fixed (small) time duration.
Here we assume that the platform has a unified trusted interface that allows clients to send requests and receive results.
In addition, a consensus protocol is also required to ensure that contents running on different computing nodes are identical. 
\footnote{We call the platform a trusted computing platform if it satisfies safety and liveness even if it is centralized. But currently, any single entity cannot satisfy these two properties, so we need a distributed system. The consensus protocol can ensure that the distributed systems can satisfy these two properties.}

A close model is the classical state machine replication (SMR) \cite{schneider1990implementing, castro1999practical}.
However, different from the SMR model, which is under the permissioned settings \cite{pass2017rethinking}, the blockchain platform allows any node to join without getting permissions from an authority, which is called permissionless. 
In the permissionless context, the set of nodes is untrustworthy, which introduces a challenging issue as that an attacker can trivially mount a ``Sybil attack''. In a Sybil attack, the attacker simply spawns lots of computing nodes and can thus easily ensure that it controls a majority of the computing nodes to achieve consensus.
To prevent Sybil attacks, a common method is to use the proof-of-work (POW) \cite{nakamoto2008bitcoin} or the proof-of-stake (POS) \cite{kiayias2017ouroboros}. 

\subsection{System Model}
In this section, we discuss the blockchain model proposed in this work in details. 
Formally, we specify the proposed blockchain model by a $4$-tuple $\Omega=(\cN,\cA,\cC,\cF)$. Here, $\cN$ is the set of all nodes in the system, $\cA$ is the set of requests, $\cC$ is the consensus protocol and $\cF$ is the set of all possible behavior errors (or failures) of the system.
This model proceeds for infinite rounds. In each round $t$, the set of available nodes and the set of requests are given.  We use $\cN_t\subseteq \cN$ and $\cA_t\subseteq \cA$ to denote the current sets of available nodes and requests, respectively.
The goal is to generate a consistent log for each node in $\cN_t$ to process the set of requests $\cA_t$.

For each round $t$, we use $T_t$ to denote the duration of round $t$ and $d_{t,i}$ to denote the confirmation time of request $i\in \cA_t$.
Then we have $T_t=\max_{i\in \cA_t}d_{t,i}$ and average confirmation time $d_t=\sum_{i\in \cA_t}d_{t,i}/|\cA_t|$.
We define the \emph{throughput} of round $t$ as the number of requests processed per second, which is equal to $\frac{|\cA_t|}{T_t}$.
Furthermore, to implement the consensus protocol $\cC$ in the proposed blockchain infrastructure, we introduce ``log'', a  combination of timestamps, to define a series of an ordered way (see Section 2.2  for more details). For any two given nodes, they will run the same request and reach the same state if and only if they have the same log. 

\subsection{System Design Goals}
\subsubsection{The Consensus}\label{sec:con}
The consensus protocol $\cC$ is the core of the distributed computing platform.
A consensus protocol refers to an algorithm for a set of interactive Turing Machine (also called nodes) to interact with other nodes. 
We start with an interactive Turing Machine $\sM$ and a set of requests $\cA$. As discussed above, a \emph{log} $b$ for $\sM$ is a combination of timestamps to define a series of ordered way.
Usually, these requests are arranged in a linear, ordered way. It can also be arranged in parallel, even in the form of directed acyclic graphs (DAG). Since every directed acyclic graph has a topological ordering, we assume that the order of the log is linear.
Formally, we have
$$
b=\{(\pi_1,t_1),(\pi_2,t_2),\ldots,(\pi_r,t_r),\ldots\}
$$
where $\pi$ denotes requests of $\cA$, and $t$ denotes times in  $\mathbb{R}_{\ge 0}$. \liu{please double check the symbol.}
A \emph{timed execution} of a machine $B=(\sM,b)$ is defined to be a sequence 
$$
\alpha= s_0,(\pi_1,t_1),s_1,(\pi_2,t_2),\ldots,(\pi_r,t_r),s_r,\ldots
$$
where the $s$ represents states of the machine $\sM$ and $s_0$ is the initial state. We require that the sequence of the successive times $t_r$ in $\alpha$ be nondecreasing.
In the consensus protocol $\cC$, each node maintains a local linear log that satisfies the following conditions.
\begin{itemize}
\item {\em Consistency:} at any point in the execution, all honest nodes have consistent logs, \ie either their logs are identical, or one node's log is a prefix of the other's.
\item {\em Synchronism:} whenever a node sees some requests in its log, the same request will appear in every other node's log within some fixed (small) amount of time.
\end{itemize}
The goal of the system is to generate a consistent log of each node in $\cN_t$ to process the set of requests $\cA_t$.

\subsubsection{Reliability and Security}
Failures and attacks can occur in many systems, especially in distributed systems consisting of heterogeneous individual computing nodes. To analyze the reliability and security of a system, it is crucial to correctly model failures and attacks faced by the system. We use $\cF$ to specify possible failures of nodes and the network, as well as the attack model.

The analysis of \emph{reliability} is based on the assumption of the failure set $\cF$. With a regular system setting, it should ensure that the system can achieve the desired performance with negligible probability of system failures.  

When malicious attacks exist, system security may be under risk. We aim to design our system against typical attacks, such as denial-of-service attacks, Sybil attacks, and Byzantine behavior of nodes. 
\liu{(If we don't want to claim that the system can handle all types of attacks, we can just specify what attacks we can handle. )}

\subsubsection{The Scalability}\label{sec:scale}
Throughput is an important system performance metric, which represents the system ability to handle requests. However, this indicator itself is not adequate.
When the system reaches its processing limit, it has to discard excess requests requested by clients, making the system much less usable.
 Therefore, we need to have another performance metric to evaluate the system limit, which is {\em scalability}. 

The term \emph{scalability} is often found in blockchain literature to indicate that the system throughput can monotonically increase when the number of computing nodes increases in the system.  
This definition is not accurate and makes it difficult to evaluate and compare different blockchain systems. 
In addition, the scalability is often multifaceted, which can represent (1) load scalability, the ability for a system to accommodate heavier or lighter loads, (2) functional scalability , the ability to enhance the system’s capabilities by adding new functions, and (3) generation scalability, the ability to scale up by using new generations of components. 

In this paper, we present a framework for analyzing the scalability of blockchain systems, which is somewhat similar to that of distributed systems \cite{jogalekar2000evaluating}.
Given a configuration $\sC$ for the system, we use $\sF_{\sC}$ to evaluate the performance and cost of the system by \liu{what does economy mean? the economic costs?} 
\begin{equation}
\sF_{\sC}=\sF(\sT,\sS,\sQ)
=\frac{\sT\cdot\sQ}{\sS}
\end{equation}
where
$\sT$ and $\sQ$ denote the throughput and the quality of the service (QoS) of the system respectively. 
$\sS$ denotes the system overall costs, including the node costs, network bandwidth costs and so on.
For a blockchain, its QoS $\sQ$ is mainly determined by the average confirmation time $d$ and the target confirmation time $\hat{d}$. In order to normalize the QoS to $(0,1)$ interval, we set $\sQ=\frac{\hat{d}}{d+\hat{d}}$. 
Then we have $Q=0$ when $d\rightarrow \infty$ and $Q=1$ when $d\rightarrow 0$.

Given an initial configuration $\sC$, 
we can scale up the configuration of the system to $\sC_k$ by a scaling factor $k$.
A \emph{policy} $\sigma$ specifies how to scale up the configuration.
For example, when the initial configuration $\sC$ has $n$ nodes, and the policy $\sigma$ scales up the number of nodes by $k$, the number of nodes in configuration $\sC_k$ is equal to $k\cdot n$.
Then, we can quantitatively calculate the scalability as 
\begin{equation}\label{equ:scable}
\psi_{\sigma}(k)=\frac{\sF_{\sC_k}}{\sF_{\sC}}
\end{equation}
If $\psi_{\sigma}(k)$ equals to $1$ or monotonically increases as $k$ increases, we say the system has \emph{perfect scalability} under the policy $\sigma$.

\begin{figure*}[tb]
\centering
\includegraphics[width=6.5in]{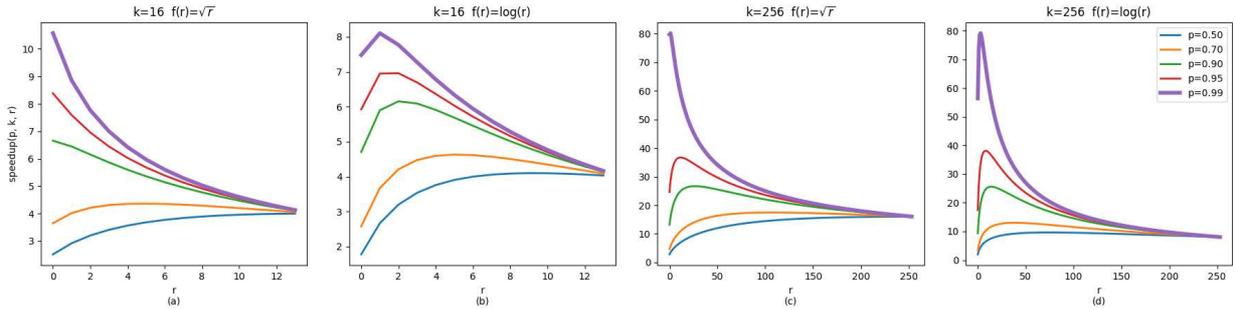}
\caption{Speedup with (a) $k=16, f(r)=\sqrt{r}$, (b) $k=16, f(r)=\log r$, (c) $k=256,f(r)=\sqrt{r}$ and (d) $k=256,f(r)=\log r$.}\label{fig:f}
\end{figure*}
\subsubsection{The Speedup}
Recall from Section \ref{sec:scale} that the scalability of a system is based on scale-up policy $\sigma$.
There are two ways to improve system performance. One is by increasing the performance of each node, and the other is by increasing the number of nodes. 
Similar to  Amdahl’s Law \cite{amdahl1967validity,hill2008amdahl} which is a formula to compute the theoretical speedup for distributed systems, we define the {\em speedup} for blockchain systems. 
Suppose that we use $k$ times the budget to build the platform, part to improve the performance of each node, and part to increase the number of nodes. Here we assume that all the nodes are identical.
We use $r$ times the budget to improve node performance, which can be increased by $f(r)$ times. 
The function $f(r)$ can be any arbitrary functions. Because of hardware limitations, usually we set $f(r)=\sqrt{r}$ or $\log r$. 
Then the number of nodes can be expanded by $\frac{k}{r}$ times.
For the request set $\cA$, assume that the proportional $p$ part can be processed in parallel, while the remaining $1-p$ parts can only be serialized (\eg $p=0$ in Bitcoin). Then the speedup ratio of the entire system can be expressed as:
\begin{equation}
\mathsf{Speedup}(p,k,r)=\frac{1}{\frac{1-p}{f(r)}+\frac{p\cdot r}{f(r)\cdot k}}
\end{equation}
and the metric $\psi_{\sigma}(k)$ can be approximated as 
$$\psi_{\sigma}(k)\approx\mathsf{Speedup}(p,k,r)/k.$$
\liu{I suggest to write scalability as another equation as it is an important performance metric. } 
As shown in Figure \ref{fig:f}, a larger $p$ value leads to a higher speedup, which agrees with our intuition. 
The results inspire us to view the entire system's performance rather than focusing on node efficiencies.
 In particular, the design of the system should focus on the support of more parallelism through architectural and protocol design.

Besides, there are also some optimizations that can improve the performance and scalability of a blockchain system. For example, use DAG instead of the linear log and use POS protocol instead of the POW protocol.
\section{The Architecture of Thinkey}
In this section, we focus on the design of a blockchain infrastructure with high scalability without weakening reliability and security.
\subsection{A Four-layer Encapsulation System Structure}
\begin{figure}[tb]
\centering
\includegraphics[width=3.0in]{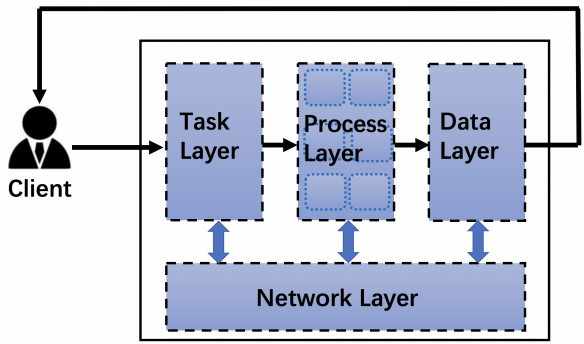}
\caption{The high-level structure of Thinkey overview. 1. Task layer: assign requests to specific processing committees.  2. Process layer: process requests and generates logs. 3. Data layer: synchronize the logs of committees to achieve consistency. 4. Network layer: provides communication between nodes.}\label{fig:p0}

\end{figure}

Based on the above requirements, we introduce the high-level structure of Thinkey, which is a four-layer structure to support more parallel executions. As shown in Figure \ref{fig:p0}, the structure has four layers. 
The first layer is the task layer, which is mainly responsible for dividing requests and nodes and assigning different requests to specific committees for processing. The second layer is the process layer, which needs to process the assigned requests and produce logs. The third layer is the data layer, where the log and request data generated by each committee need to be aggregated according to a specific coding method to form a unified log. The fourth layer is the network layer, which is the basic layer to establish connections between nodes and provide communication. Next, we will discuss each layer in details.

{\em Task layer.}
All requests are first sent to the task layer, where they will be divided and assigned to different committees for parallel processing. As not all requests can be processed in parallel, they need to be divided according to their types. 
In addition, all active nodes are registered at the task layer. These nodes are divided into committees and allocated to process different requests in a random way. 
We need to ensure that each committee is trusted, \ie the proportion of malicious nodes inside each committee will not exceed a certain threshold set by the system.

{\em Process layer.}
The process layer has a set of committees, where each committee contains a set of nodes.
When a committee receives a given request, it needs to process the request, reach consensus, and generate a log. As each committee is guaranteed to be trustful at the task layer, this layer only needs to consider how to reach consensus in a committee as soon as possible. 

{\em Data layer.}
The goal of the system is that each node produces a consistent log. Therefore, an aggregation algorithm is required to integrate all the logs generated by the nodes in a committee to reach a unified log. An encoding method is also required to reduce the storage of each node.
In addition, as nodes should join and leave different committees from time to time, it is essential to have the corresponding data from the data layer be synchronized.
 
{\em Network layer.}
This layer provides the foundation of the entire system by enabling communications among different computing nodes. At the same time, within the network layer, we can build a multi-layer network and establish a layer of consensus network for each committee. 

As a summary, in our architecture, a request generated by a client first passes through the task layer. After the classification at the task layer, the request is sent to a committee at the process layer. The committee processes the requests and sends the results and logs to the data layer. The data layer aggregates all the logs and returns the results back to the client.


\subsection{A Double-layer Chain Architecture}
\begin{figure}[tb]
\centering
\includegraphics[width=3.1in]{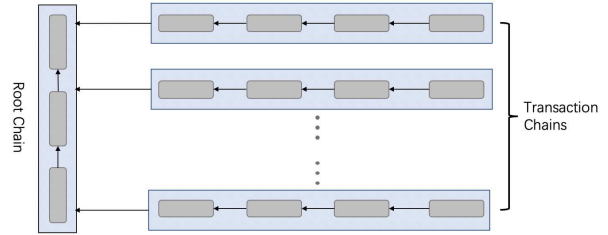}
\caption{The basic double-layer architecture of Thinkey. The root chain records metadata and digests of comfirmed blocks of each transaction chain, produces random seeds and records the election results.}\label{fig:struc1}
\end{figure}

From an implementation perspective,
our chain structure is based on a double-layer chain architecture consisting of
a \emph{root chain} and \emph{transaction chains}.
Each (root or transaction) chain is an individual blockchain with
its own state. 
As illustrated in Figure \ref{fig:struc1}, transaction chains (\ie the horizontal chains) are anchored to the root chain (the vertical chain).

The root chain acts as the leader and coordinator of the entire system. In particular, it
 serves as an entry point and source of trust to the
transaction chains; records metadata and digests of confirmed
blocks of each transaction chain; produces random seeds used in committee elections (of all
chains); and records the election results.

The workload of the system is shared by the transaction chains.
The cross-chain operations are handled using a message-driven
protocol based on the actor model (see Section~\ref{sec:actor}).

All nodes in the system maintain the state of the root chain.
By updating and verifying blocks of the root chain, a node can verify 
(any part of) a block of a transaction chain that has been included
in the root chain. This structure is advantageous in the following
aspects.
\begin{itemize}
	\item To join the system, a node only need to obtain the current state of the root chain either from a trusted source or by reconstructing from the genesis block, which is a small workload compared to synchronizing all the data of the entire system.
	\item The consensus of each (root or transaction) chain is carried out independently and in parallel, dramatically reducing the usage of network and processing power.
	\item The root chain can serve as a coordinator of the system, which provides synchrony across transaction chains and allows system topology to be dynamic.
	\item A node can verify a transaction initiated from another transaction chain, using the digest in the root chain and a Merkle proof. Therefore, a block producer of a transaction chain does not need any information from other transaction chains to process inter-chain transactions.
\end{itemize}

\begin{figure}[tb]
\centering
\includegraphics[width=3.1in]{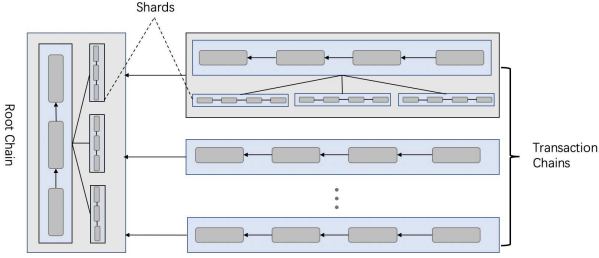}
\caption{Sharding for root chain and transaction chain in the double-chain architecture of Thinkey}\label{fig:struc3}
\end{figure}

Furthermore, the root chain and transactions chains can perform deeper sharding (see Figure \ref{fig:struc3}).
Here, the root chain and transaction chains define rules and make a balanced distribution of transactions to its sub-chains. Then it only needs to process a small number of transactions and record the digest of each sub-chain.

To sum up, such a layered fragmentation structure has great flexibility and scalability and can be dynamically adjusted, so that each chain does not become a performance bottleneck of the entire network. In addition, as the number of chains increases, the throughput of the entire system increases linearly without generating too many redundant messages.


\section{Our Protocol}

In this section, we present our protocol in more details.
In the basic double-layer architecture (see Figure \ref{fig:struc1}),
let $\iC_R$ denote the root chain and let $\iC_1, \iC_2, \ldots, \iC_n$ denote the $n$ transaction chains.

\subsection{Committee Selection}

To resist the Sybil attack against the permissionless system, we use a proof-of-stake based
election algorithm. The following security properties are desirable for the selection algorithm.
\begin{itemize}
\item Assuming $\delta_0$ honest majority (of stake) among all participants, at least
$\delta_1$-proportion of the committee members elected in each election are honest except with negligible probability.
Furthermore, the algorithm should be \emph{fair} in the sense
that the probability of each participant being chosen is (roughly)
proportional to the amount of stake committed by the participant.

\item The committee members should be fluid and unpredictable so that the adversary cannot attack the system by corrupting
committee members (assuming corruption takes longer than the life span of a committee).
\end{itemize}
In Thinkey, we achieve the above properties through the following processes. First, before the election, since only the root chain is listened to by all nodes and the chains are not synchronized, a transaction chain has to put a signal on the root chain when it needs to elect the next committee. The election will use the next seed after the signal shows up on the root chain.

Next, the election process is managed by the root chain. 
In particular, the randomness used in the election is provided by the random seed generated in the root chain $\iC_R$. The committee of the root chain generates a random seed periodically, which is included in a block of $\iC_R$. Our current implementation generates random seeds using DFINITY’s random beacon protocol \cite{hanke2018dfinity}, based on BLS  threshold signature.

In addition, nodes willing to
participate in the consensus need to register on the root
chain by sending a special type of transaction. The transaction also
specifies the amount of the stake, which will be transferred to
a specific stake account and be frozen until the node quits
and withdraws the stake.
These nodes then monitor the election signals in the root chain. When a node sees an election signal and the next seed, with its secret key, it can compute whether it is elected to the next
committee of the chain that sent the signal.

Once the committee selection is completed, nodes elected to a transaction chain will join
the network of the chain and start receiving the blocks
of the chain. They also start to synchronize the state
of the transaction chain. The received blocks and the
state can all be verified using the digests on the root chain.
Each node sends an ``elected'' transaction
containing its public key and a proof that it has been elected,
so that the current committee and the other elected members
are notified.

The elected nodes will set up a small \emph{consensus network}
that will be used for the communication within the committee.
Having a dedicated network reduces the delay and bandwidth
consumption of both point-to-point communications and broadcasts
among committee members.
On the other hand, if not set up properly, the network may
be less stable and more vulnerable to attacks.
It needs to be ensured that the network topology is robust 
and node information is exchanged securely using encryption.

By the time the next epoch begins, the new committee should
have generated the keys for threshold signature (the public
key should have been written to a block); have synchronized and updated the current state of the chain; and have established connections in the new consensus network.

\subsection{Consensus in Committees}
We assume a partially synchronous communication model within the committee, for which there exist efficient Byzantine-fault-tolerant algorithms.
Here we present a variant of PBFT tailored for this setting.

\begin{figure}[tb]
\centering
\includegraphics[width=3.0in]{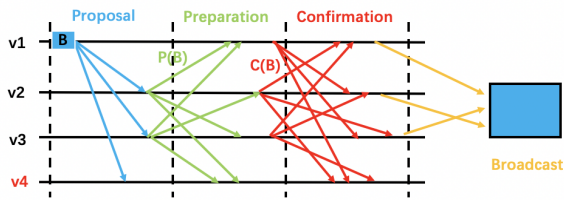}
\caption{Overview of committee consensus.}\label{fig:cc}
\end{figure}
The execution of a node can be divided into \emph{rounds}. Each round consists of three stages: proposal, preparation, and confirmation (see Figure \ref{fig:cc}). 
The state transitions are event-driven. In order to preserve system liveness in the presence of network failures or malicious attacks, a timeout may be triggered by the local clock.

\begin{enumerate}
\setlength{\itemsep}{0pt}\setlength{\parsep}{0pt}\setlength{\parskip}{0pt}
\item {\em Proposal stage}: The leader of the committee broadcasts the proposed block to other committee members.
\item {\em Preparation stage}: Upon receiving the proposed block, each committee member broadcasts a message containing its signature of the block. If a timeout is triggered before the proposed block is received, the committee member signs and broadcasts a special message (which indicates the leader is faulty) to other committee members.
\item {\em Confirmation stage}: At the end of the preparation stage, each committee member signs and broadcasts a package of the signatures received in the preparation stage.
Signature aggregation can be used to significantly reduce the size of messages in the confirmation stage.
\end{enumerate}
 Based on the messages received in the confirmation stage, each committee member can decide whether agreement on the block has been reached, and broadcasts the agreed block or an empty block together with a proof of its decision.

{\em Malicious nodes punishment.} In the case when an explicitly misbehaving node is detected (\eg a node that sends different messages to different nodes at the same stage), the round will abort by outputting an empty block. However, the misbehaving node will be economically punished by a large amount, making such attacks unsustainable.

 {\em Optimization.} In the case when the number of signatures received at the preparation stage implies that most honest committee members have received the same proposed block, a committee member may reach an ``early consensus'': the member can output the block before the confirmation stage with the signatures being the proof of agreement (of a different form compared to a regular agreement). Note that the node still needs to participate in the confirmation stage.

\subsection{Cross-Chain Message and Verification}\label{sec:crosschain}
In a multi-chain system, a cross-chain operation is unavoidable. Each chain needs to process some messages which are generated by other chains. 
There are two kinds of cross chain messages in our system. The first one is the message $m_i$ from $\iC_i$ to $\iC_R$ which contains the digest of a block of $\iC_i$. The message $m_i$ is for the final confirmation of the block of $\iC_i$ and there is only one such message for each block.
The second one is the the outer relay message $m_{i,j}$ from $\iC_i$ to $\iC_j$ (see Section \ref{sec:actor} for more details). Before $m_{i,j}$ is sent to $\iC_j$, it is recorded on the block of $\iC_i$.

Before the deal with these messages, we should verify them first. There are two approaches to verify a message generated by a chain: (1) verify the signature; (2) verify the message hash. 
Both approaches are useful, depending on the type of messages.
For the message $m_i$ from $\iC_i$ to $\iC_R$, it attaches a signature of the committee members of $\iC_i$ that can be used to verify authenticity.
 Since the committee member of $\iC_i$ is recorded in $\iC_R$, each node in $\iC_R$ has the public keys of current committee members of $\iC_i$ and can verify the signature of $m_{i}$.

For the message $m_{i,j}$ from $\iC_i$ to $\iC_j$, it comes with proof $\pi_{i,j}$ that can be proved. 
Before $m_{i,j}$ is sent to $\iC_j$, it is recorded on the block of $\iC_i$. On the block of $\iC_i$, there is a Merkel tree $\rT_i$ for all outer relay messages.
The proof $\pi_{i,j}$ refers to the hash values of all sibling nodes on the path from the Merkle tree root to its entry. 
The root hash of $\rT_i$ is included in the digest of $\iC_i$. Since the digest of the block of  $\iC_i$ is recorded in the root chain $\iC_R$, and each node in $\iC_j$ is also a node in $\iC_R$, each node in $\iC_j$ can obtain the root hash of $\rT_i$ and verifies $m_{i,j}$ by the proof $\pi_{i,j}$.

The reason for not using the signature to verify $m_{i,j}$ is to prevent the situation where the committee members of $\iC_i$ are not completely reliable. 
The message generated by this block will only take effect if its block is finally confirmed on $\iC$.
However, for a cross-shard message of the same transaction chain $\iC_i$ (see Figure \ref{fig:struc3}), its verification can be faster, just wait until the digest is recorded on its transaction chain $\iC_i$, not the root chain $\iC$.

\subsection{Network Layer}\label{sec:network}
The decentralized nature of P2P networks makes them an ideal solution for the underlying communications of blockchain systems. As the network part often becomes the bottleneck of high throughput blockchains, P2P technology has attracted more and more attention as a potential technology for blockchain networks.  
In blockchain, as a large number of data broadcasting operations are required, efficiency and redundancy are the key issues that all blockchain P2P network design must pay attention to. 

Conventional blockchain P2P networks (such as Bitcoin and Ethereum) are often based on unstructured design ideas. When broadcasting, a partial-broadcasting mechanism is adapted to alleviate the message redundancy problem. For example, when broadcasting a large message $M$, node $P$ randomly selects some nodes in its connection list for full transmission. For the remaining nodes (\eg $Q$), only the profile of $M$ (\eg the hash of $M$) is transmitted. Only when $Q$ needs the message $M$ and cannot find it in its own database, it will pull the message from $P$. In the experiment part, we have proved through comparison of several sets of parameters that the design can significantly reduce the redundancy of message transmissions.
\begin{figure}[tp]
\centering
\includegraphics[width=3.0in]{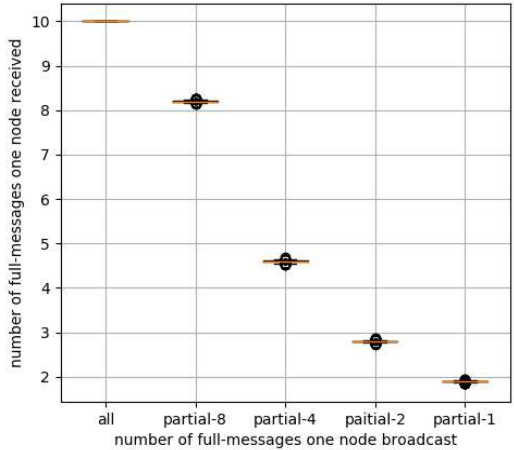}
\caption{Data redundancy.}\label{fig:e1}
\end{figure}

However, this solution still has its limitations. For example, the requirements for higher throughput and point-to-point transmission will become inadequate under a structured P2P network. In such cases, the structured network (such as DHT) is a solution that can be used for further optimization. In a structured P2P network, nodes follow some rules during the networking phase (\eg Chord \cite{stoica2001chord}, Kademlia \cite{maymounkov2002kademlia}). This provides directional guidance for data transfer and can greatly reduce the number of useless transfers. However, in blockchain, especially in public chain projects, there are assumptions that nodes can join and exit arbitrarily. Networks with frequent changes will incur huge costs for structure maintenance. Therefore, one of the challenges in adopting structured P2P networks is to deal with the instability and uncertainty of complex networks. In Thinkey, we use a combination of structured + unstructured methods. Some scenarios (such as point-to-point transmission) use a structured P2P network approach to reduce redundancy, and to improve efficiency, while having an unstructured transmission as a guarantee of stability.

Figure \ref{fig:e1} shows the impact of the partial-broadcasting mechanism in reducing message redundancy. We assume that each node is randomly connected to 10 nodes. The horizontal axis shows the number of complete information received by each node when the node sends complete information to all, $8$, $4$, $2$, and $1$ neighbors. For each case, we count the redundancy of the network when it expands from 100 nodes to 1000 nodes. It can be seen that (1) the system redundancy can be significantly reduced when the number of full message broadcast decreases. When a node broadcasts a complete message to only one neighbor, each node will only receive the same complete information less than twice. (2) When the number of neighbors connected to each node is constant, the amount of redundancy has nothing to do with the size of the network.


\section{A New Model of Account in the Multi-Chain Structure}\label{sec:actor}
For the multi-chain system, the current account models (\eg UTXO or Ethereum account) for the single-chain system have not adapted to meet the new realities, especially when dealing with a large number of cross-chain operations. 
Based on the theoretical of Actor Model \cite{hewitt1973universal}, we propose a new account model, which allows us to implement complex logic on the multi-chain system in an asynchronous and lock-free manner.
In this model, we decouple a transaction which involves a set of accounts into multiple steps in the form of messages. Each message is received by a unique recipient and will be executed by the corresponding chain. 
All messages will be executed and will eventually achieve the transaction correctly.

\subsection{Actor Model}
Actor Model was introduceted by Hewitt $\etal$ \cite{hewitt1973universal} in 1973, which is a conceptual model to deal with concurrent computation. 
The model has been used both as a framework to understand concurrency and as the theoretical basis for several practical implementations of concurrent systems.
For example, the electronic mail system, web services and objects with locks in Java. See \cite{hewitt2010actor} for more details.

\begin{figure}[tb]
\centering
\includegraphics[width=3.0in]{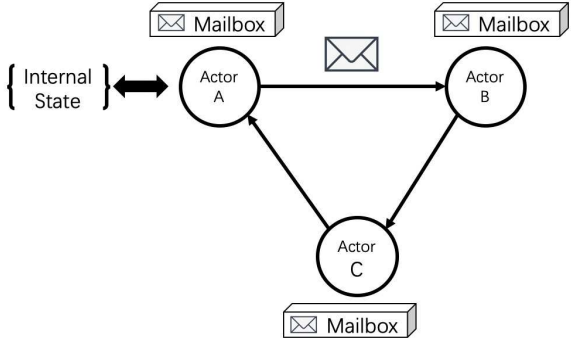}
\caption{Actor Model}\label{fig:actor}
\end{figure}

An \emph{actor} is the universal primitives of concurrent digital computation. When receiving a message, the actor can concurrently:
\begin{enumerate}
\item send a finite number of messages to other actors; 
\item create a finite number of new actors; 
\item designate the behavior to be used when the next message is received. 
\end{enumerate}
There is no assumed sequence to the above actions and they could be carried out in parallel.
Each actor has a mailbox to receive messages from other actors (see Figure \ref{fig:actor}).
Messages are sent asynchronously to an actor and each actor processes messages sequentially with no restriction on message arrival order. 
Multiple actors can run at the same time. It is worth noting that actors are completely isolated from each other and they will never share memory. Each actor maintains a private state that can never be changed directly by another actor.

\subsection{Thinkey Account}
We adopt Actor Model as the basic framework for our account model.
In Thinkey, the account structure mainly contains the following field pieces of information:

\hangafter=1 \setlength{\hangindent}{2.5em}
{ Address}: The unique identifier of the blockchain account.

\hangafter=1 \setlength{\hangindent}{2.5em}
{ Balance}: The account's current balance.

\hangafter=1\setlength{\hangindent}{2.5em}
{ Nonce}: A scalar value equal to the number of externally messages sent from this address.

\hangafter=1 \setlength{\hangindent}{2.5em}
{ Code}: The programmatic logic for processing messages. 

\hangafter=1 \setlength{\hangindent}{2.5em}
{\em Storage}: The internal state of the account which can be empty.

\noindent Each account is controlled by a private key. 
In the code, the account defines its own methods for the messages it received. It is allowed to send a message to other accounts, to create a new account and to modify the internal state. 
For some specfic messages, there are general processing methods which are identical for each account (\eg ``$\mathsf{tran}$'' and ``$\mathsf{add}$'' in Section \ref{sec:payment}). 
Each account also can customize the methods for other messages. 

There are two types of message: \emph{external message} and \emph{relay message}. 
The external message is created by an account who signs it with its private key. 
The relay message is produced by an account who executes the send command in the process of execution, which is somewhat similar to the message in Ethereum.
The biggest difference is that the relay message execution is asynchronous in our model and synchronous in Ethereum.
Thus these messages in our model support cross-chain propagation.
 A message in Thinkey mainly contains the following field pieces of information:

\hangafter=1 \setlength{\hangindent}{2.5em}
{ From}: The address of the messages' sender. 

\hangafter=1\setlength{\hangindent}{2.5em}
{ To}: The address of the recipient of the message.

\hangafter=1\setlength{\hangindent}{2.5em}
{ Nonce}: A scalar value equal to the number of external messages sent by the sender or, for relay messages, nil.

\hangafter=1 \setlength{\hangindent}{2.5em}
{ Input}: A tuple specifying the input data of the message call.

\hangafter=1 \setlength{\hangindent}{2.5em}
{ Verification data}: A signature identifying the sender for externally message, or a proof for the relay message. 

\noindent For an external message, it can be verified by the signature and nonce. For a relay message, it can be verified by the proof (see Section ).

\subsection{Thinkey Block}
\begin{figure}[t]
\centering
\includegraphics[width=3.0in]{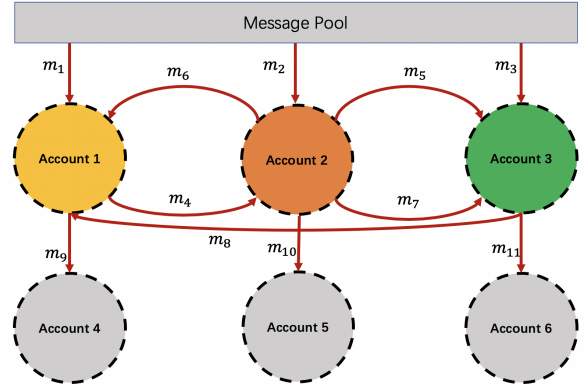}
\caption{An example of processing for the accounts in our model}\label{fig:complicated}
\end{figure}

For each block on a chain $\iC$, there is three status of a message on the block. 
\begin{enumerate}
\item Input messages. These messages are currently not confirmed and the recipients of them are in $\iC$. They can be external messages or relay messages produced by other chains.
\item Inter relay messages. These are the messages generated during the execution of the entire block, and their receivers are on the same chain $\iC$. Thus they are confirmed on the block.
\item Outer relay messages. These are the messages generated during the execution of the entire block, and their receivers are on other chains. These messages will be confirmed by other chains.
\end{enumerate}
We describe an example shown in Figure \ref{fig:complicated}. There are three accounts $1,2,3$ on the chain $\iC$ and three accounts $4,5,6$ on other chains. In particular, there are three input messages (\ie $m_1,m_2,m_3$), received by the accounts $1,2,$ and $3$ respectively. 

For each account $i$, we use $\sigma_i$ to denote the order in which the messages are processed and the new messages are generated. Then, we have
\begin{align*}
\sigma_1 &=(m_1:m_4\ |\ m_8\ |\ m_6:m_9)\\
\sigma_2 &=(m_2:m_5,m_6 \ |\ m_4:m_7,m_{10})\\
\sigma_3 &=(m_5: m_{11}\ |\ m_3:m_8\ |\ m_7)
\end{align*}

Thus $\sigma_1$ indicates that account $1$ processes message $m_1$, $m_8$ and $m_6$ in order. An inter relay message $m_4$ is generated by excuting $m_1$ and an outer relay message $m_9$ is generated by excuting $m_6$.
Then we have inter relay messages set $\{m_4,m_5,\ldots,m_8\}$ and outer relay messages set $\{m_9,m_{10},m_{11}\}$.
 
\begin{figure}[tb]
\centering
\includegraphics[width=3.0in]{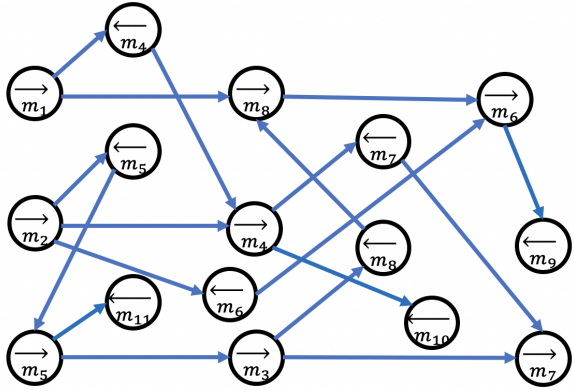}
\caption{The direct graph $G_{\sigma}$ of the processing message order $\sigma=(\sigma_1,\sigma_2,\sigma_3)$ }\label{fig:direct}
\end{figure}

On receiving a block proposed by a leader, a committee member needs to verify the block to defend against a malicious leader. In our system, a node verifies three parts: 
\begin{enumerate}
\item Validity of each input message, \ie  $\{m_1,m_2,m_3\}$.
\item Validity of the proceesing for each account, \ie  $\{\sigma_1,\sigma_2,\sigma_3\}$. They can be verified independently and in parallel.
\item Validity the processing message order, \ie the order $\sigma=(\sigma_1,\sigma_2,\sigma_3)$.
\end{enumerate}
To verify the processing message order $\sigma$, we build a direct graph $G_{\sigma}$ below.
For two events $e_1,$ and $e_2$, the expression of $e_1\rightarrow e_2$ indicates that the time of $e_1$ precedes the time of $e_2$ in the frame of reference for  every relativistic observer.
Let $\ra{m_i}$ denote the event that an account receives a message $m_i$, and $\la{m_i}$ denote the event that an account sends a message $m_i$.
 Then we have $\la{m_i}\rightarrow \ra{m_i}$. 
For each $\sigma_i$, for example $i=1$, we have $\ra{m_1}\rightarrow\ra{m_8}\rightarrow\ra{m_6}$.
Based on these relationships from $\sigma$, we can build a directed graph $G_{\sigma}$ (see Figure \ref{fig:direct}).

\begin{theorem}
The processing message order $\sigma=(\sigma_1,\sigma_2, \sigma_3)$ is valid if and only if the dircet graph $G_{\sigma}$ does not have a circle.
\end{theorem}

\subsection{A Payment Application}\label{sec:payment}
\begin{figure}[tb]
\centering
\includegraphics[width=3.0in]{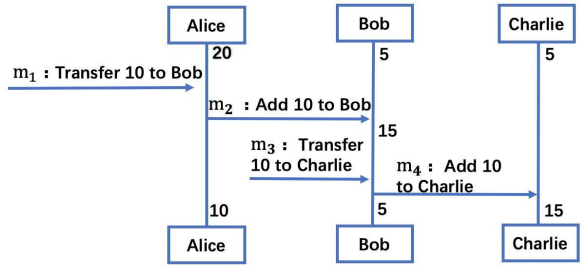}
\caption{A transfer example}\label{fig:simple}
\end{figure}

In this section, we present a payment process, such as Bitcoin, as an example of the Thinkey's potential applications. We define an account with a local state as $\mathsf{balance}$ (see below). There are two kinds of messages: ``$\mathsf{tran}$'' and ``$\mathsf{add}$'', which represent withdraw and deposit respectively.  
A simplified sample code written in the Elixir language \cite{thomas2018programming} that describe the payment process is shown in Algorithm \ref{alg:pay}.
\begin{algorithm}
\caption{The account module for payment application}
\begin{algorithmic}[1]
\STATE balance 
\WHILE{receive message $m$}
\IF{$m=\{:\text{tran, address, bill}\}$}
\IF{balance $\ge$ bill}
\STATE balance $=$ balance $-$ bill
\STATE send $(\text{address},\{:\text{add, bill}\})$
\ENDIF
\ELSIF{$m=\{:\text{add, bill}\}$}
\STATE balance $=$ balance $+$ bill
\ENDIF
\ENDWHILE
\end{algorithmic}
\label{alg:pay}
\end{algorithm}
As shown in Figure \ref{fig:simple}, we use an example to demonstrate the process. 
Suppose there are three accounts Alice, Bob, and Charlie with balance $20,5,10$ respectively.
There are two messages $m_1$ and $m_3$, which are signed by Alice and Bob respectively. Message $m_1$ is a transfer with $10$ from Alice to Bob and message $m_3$ is a transfer with $10$ from Bob to Charlie. Then as we set on the code, message $m_1$ is processed on Alice's chain as follows.
\begin{enumerate}
\setlength{\itemsep}{0pt}\setlength{\parsep}{0pt}\setlength{\parskip}{0pt}
\item An unconfirmed message $m_1$ is picked up by the committee members in Alice's chain if the signature is validated when the members construct a new block.
\item When the balance of Alice is not less than the transfer amount (\ie satisfies Line $4$ in Algorithm \ref{alg:pay}), and Alice and Bob belong to the same chain, the committee members records the message processing procedure $\sigma_a=(m_1:m_2)$ on the block. It means that processing $m_1$ produces $m_2$ (Line $6$ in Algorithm \ref{alg:pay}). Meanwhile, the message $m_2$ is marked as an inter-relay message and executed on the block. In this case, the balance of Alice is deducted by $10$ (Line $5$ in Algorithm \ref{alg:pay}) and the balance of Bob is added by $10$ (Line $9$ in Algorithm \ref{alg:pay}). 
\item When Line $5$ in Algorithm is satisfied, and Alice and Bob are from different chains,  $\sigma_a=(m_1:m_2)$ is also recorded on the block. Meanwhile, the committee members mark the $m_2$ as outer relay message on the block and send $m_2$ to the destination chain, \ie Bob's chain. 
In this case, the withdraw operation for Alice is executed (Line $5$ in Algorithm \ref{alg:pay}).
The outer relay message $m_2$ will be excuted in the same way at Bob's chain.
\item When the balance of Alice is less than the transfer amount (\ie Line $4$ in Algorithm \ref{alg:pay} is not satisfied), $\sigma_a'=(m_1:\varnothing)$ is recorded on the block. In this case, the balances of Alice and Bob do not change. 
\end{enumerate}

For each account $i$, we use $\sigma_i$ to denote the order in which the messages are processed and the new messages are generated.
Thus, when these three accounts are on a same chain, we have $\sigma_a=(m_1:m_2),\sigma_b=(m_2 \ |\ m_3:m_4),\sigma_c=(m_4)$ as shown in Figure \ref{fig:simple}, and record the  message processing procedures $\sigma=(\sigma_a,\sigma_b,\sigma_c)$ in the block. Then nodes can get the same state through $\sigma$. The only thing to do is to reach a consensus on $\sigma$ (block) because the reception order of messages can affect behaviors. 
For example, the order in which the processing messages $m_2$ and $m_3$ for Bob can be reversed to get another valid message processing procedures $\sigma'=((m_1:m_2),(m_3\ |\ m_2),\varnothing )$. 
When there three accounts are on different chains, the message processing procedures $\sigma_a$, $\sigma_b$ and $\sigma_c$ are recorded on corresponding chains respectively, which complete these two transfers. 

We remark that the transfer of balance is confirmed at the first stage. As long as the message $m_1$ is included in the block during the first phase, Bob can be certain that the transfer has been confirmed. Bob can just wait for the settlement. And the value being transferred can be used. Through this process, cross-chain transfers can achieve the same speed of confirmation as intra-chain transfers.

\subsection{CryptoKitties Application }\label{sec:kitty}

In the previous section, we present a payment application. Moreover, our model can support any complex logic.
We take CryptoKitties as another example, which is a popular application in Ethereum and attracts a lot of people. 
CryptoKitties is an electronic pet game based on the Ethereum smart contract. In the game, players can buy and sell.
However, the kitty auctions can block the Ethereum network, which undermines the system's availability and significantly increases the usage costs.
In our account model, we propose a multi-chain solution to solve network congestions.
\begin{figure}[tb]
\centering
\includegraphics[width=3.0in]{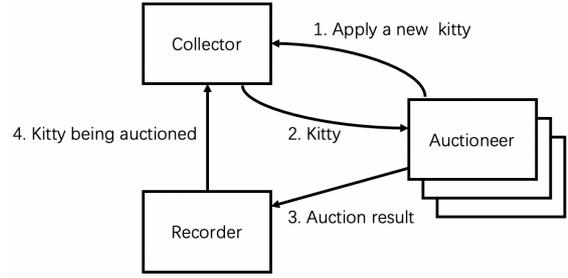}
\caption{A multi-chain solution for CryptoKitties}\label{fig:kitty}
\end{figure}

In particular, our program involves three types of accounts: a Collector, multiple Auctioneers, and a Recorder as shown in Figure \ref{fig:kitty}. The Collector is responsible for collecting all auction requests from a seller. The Auctioneers will auction the kitties. And the Recorder is responsible for counting auction results.

Processing is kicked off by an Auctioneer requesting a kitty from the Collector. When the Auctioneer receives the kitty, it can auction this kitty and sends the auction result to the Recorder.   
Finally, the Recorder lets the Collector know whether the kitty has been auctioned.
The auctioneer can be placed on different chains. The workloads of the auctions can be partitioned in multiple independent and parallel chains.

It is possible that a single kitty attracts a committee of people to bid, which is called ``single kitty hotspot'' issue.
In this case, the auction of the hot kitty can be held by different auctioneers. The bidder can choose any auctioneer to bid.
 Finally, all the auction results will be summarized and sent to the Recorder. According to the rules, the Recorder selects a bid winner for the hot kitty.

\subsection{Optimization and Discussions}
We use optimization to reduce the cost of communication and the storage of accounts. The first optimization aims to avoid duplicated storage. For some specific messages, there are general processing methods that are identical for each account. These public methods are designed by the system and are not required to be designed by each account.
 The second optimization reduces the cost of communication by merging a bunch of messages of the same type. For example, if there are 10 ``$\mathsf{add}$'' messages, which are sent to the same account, they can be combined into one ``$\mathsf{add}$'' message.

For the ``single account hotspot'' issue, where a single account is involved in a great number of messages, it can be easily resolved by the co-design of applications at the upper layer, which is similar to the ``single kitty hotspot'' issue as discussed in Section \ref{sec:kitty}.
The popular accounts can be placed on multiple chains.

\section{System Analysis and Experiment}
In this section, we will analyze the security, the performance, and complexity of Thinkey.
\subsection{Security}
Let $N$ be the number of nodes, $n$ be the expected number of nodes in a committee, $m$ be the number of committees. 
The number of malicious nodes is $\lambda N$.
We say a committee is {\em failed} when more than $\rho$ of the nodes in the committee are malicious nodes.
Without loss of generality, we set $N=n\cdot m$. 
Suppose there is a completely random oracle $\cO: [N]\rightarrow [m]$. 
Fix a committee $i$, let $A_i$ be the event that the fraction of malicious nodes in the committee $i$ $>\rho$.
Then for each $i\in [m]$, we have
\begin{equation}
\Pr[A_i]
=\sum_{x=\rho n+1}^{n}\frac{{\lambda N\choose x}\cdot{(1-\lambda)N \choose n-x}}{{N \choose n}}.
\end{equation}
By union bound, we have
\begin{equation}
\Pr\left[\cup_{i\in [m]}A_i\right] \le m\cdot \Pr[A_i].
\end{equation}
We can ensure that the probability of the event occurring is negligible for an adequate parameter setting. 

\subsection{Experiment}\label{sec:exper}
In the experiment part, we simulate a blockchain system including 400 nodes to analyze the performance of our protocol. Each node works on a single core CPU with 2G RAM. Our system is implemented using Go. 

\begin{figure}[tb]
\centering
\includegraphics[width=3.0in]{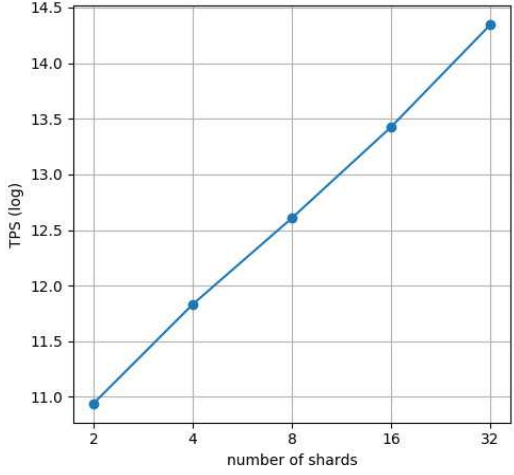}
\caption{The relationship between TPS and the number of shards.}\label{fig:e2}
\end{figure}

\subsubsection{Scalability}
In the first experiment, we analyze the relationship between the throughput per second (TPS) and the number of shards in this system. 
Here, all the accounts involved in the transaction are randomly generated.
As shown in Figure \ref{fig:e2}, 
the linear increasing curve shows that Thinkey achieves our design goal on the scalability through the proposed sharding scheme. 
That is, when the number of chains (shards) increases, the TPS can also increase linearly, instead of stagnation of single-chain system (\ie Bitcoin). 

\begin{figure}[tb]
\centering
\includegraphics[width=3.0in]{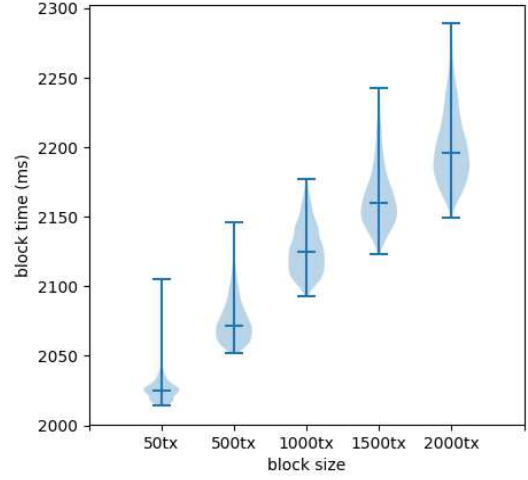}
\caption{The relationship between the block time and the block size.}\label{fig:e4}
\end{figure}

\subsubsection{Confirmation Latency}
The confirmation latency is completely related to the block time (\ie the time interval between two adjacent blocks).
In the second experiment, we analyze the relationship between the block time and block size (\ie the number of messages in the block).
As shown in Figure \ref{fig:e4}, when the block size is fixed, 
the distribution of the block time roughly follows Gaussian distribution with small fluctuations.
In addition, when the block size increases, the block time also increases monotonically. 
This result indicates that the TPS cannot be increased directly by increasing the block size. 
Comparing the result in the previous section, the best way to scale out the TPS is to use the sharding scheme.

\begin{figure}[tb]
\centering
\includegraphics[width=3.0in]{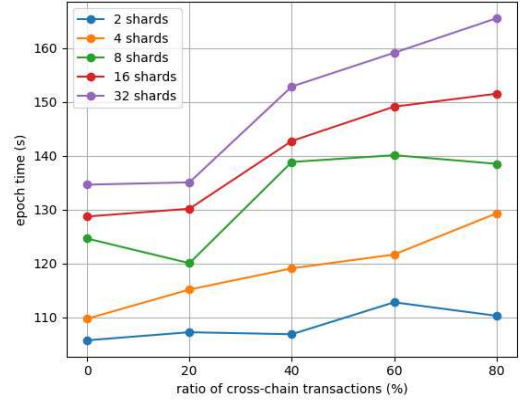}
\caption{The relationship between the epoch time, the number of shards and the ratio of cross-chain transactions.}\label{fig:e5}
\end{figure}

\subsubsection{Cross-chain Overhead}
In a multi-chain system, a cross-chain operation is unavoidable. 
Especially when the number of chains is large, the ratio of cross-chain transactions (\ie percentage of cross-chain transactions among all transactions) will be high.
In the third experiment, we analyze the relationship between the epoch time and the ratio of cross-chain transactions. 
We generate different numbers of shards, \ie setting it to $2,4,8,16$ and $32$.
 As shown in Figure \ref{fig:e5}, the epoch time increases when either the number of shards or the ratio of cross-chain transactions increases.
However, when the number of shards is fixed, the smooth curve shows that the overhead of the cross-chain transaction is small, which achievers our design goal on handling cross-chain transactions through a new model of accounts.
That is, processing cross-chain transactions in an efficient and scalable way.

\section{Conclusion}
We proposed and implemented Thinkey, a distributed blockchain architecture that enables efficient, secure and highly scalable transaction handling. 
We proposed a four-layer encapsulation system structure and a double-layer chain architecture.
 In particular, the four-layer structure enables the abstraction and separation of diverse system functionalities and allows each layer to focus on its own functionality. 
The double-layer chain architecture can make the system scale out linearly by partitioning the workload of the system.
Additionally, committee selection and cross-chain verification are key protocols proposed by our system, ensuring the efficiency and security of systems.
We also proposed a novel account model to support any complex logic in a multi-chain structure.
In our experiment, we demonstrate that our system delivers $10000+$ throughput.


{\footnotesize \bibliographystyle{acm}
\bibliography{0_main}


\end{document}